\documentclass[sigconf]{acmart}
\usepackage{balance}

\usepackage{color}
\usepackage{graphicx}

\usepackage{array}
\usepackage{multirow}
\usepackage{float}
\usepackage{subcaption}
\usepackage{footnote}
\makesavenoteenv{tabular}
\makesavenoteenv{table}
\usepackage{makecell}

\usepackage{amsmath,amsfonts,bm}
\usepackage[super]{nth}

\usepackage{url}
\usepackage{hyperref}
\hypersetup{colorlinks=True, urlcolor=black}

\usepackage[ruled,vlined]{algorithm2e}
\usepackage{enumitem}
\def\ours{PALLE}

\copyrightyear{2025}
\acmYear{2025}
\setcopyright{acmlicensed}\acmConference[MM '25]{Proceedings of the 33rd ACM International Conference on Multimedia}{October 27--31, 2025}{Dublin, Ireland}
\acmBooktitle{Proceedings of the 33rd ACM International Conference on Multimedia (MM '25), October 27--31, 2025, Dublin, Ireland}
\acmDOI{10.1145/3746027.3754745}
\acmISBN{979-8-4007-2035-2/2025/10}

\begin{document}

\title{Pseudo-Autoregressive Neural Codec Language Models for Efficient Zero-Shot Text-to-Speech Synthesis}

\author{Yifan Yang}
\authornote{Work done during an internship at Microsoft.}
\orcid{0009-0003-0588-1812}
\affiliation{
  \institution{Shanghai Jiao Tong University}
  \city{Shanghai}
  \country{China}
}

\author{Shujie Liu}
\orcid{0009-0008-2599-6752}
\affiliation{
  \institution{Microsoft Corporation}
  \city{}
  \country{Hong Kong}
}

\author{Jinyu Li}
\orcid{0000-0002-1089-9748}
\affiliation{
  \institution{Microsoft Corporation}
  \city{Redmond}
  \country{USA}
}

\author{Yuxuan Hu}
\orcid{0009-0004-5688-7488}
\affiliation{
  \institution{Microsoft Corporation}
  \city{Redmond}
  \country{USA}
}

\author{Haibin Wu}
\orcid{0000-0001-7166-5534}
\affiliation{
  \institution{Microsoft Corporation}
  \city{Redmond}
  \country{USA}
}

\author{Hui Wang}
\orcid{0009-0003-8057-4644}
\affiliation{
  \institution{Microsoft Corporation}
  \city{Beijing}
  \country{China}
}

\author{Jianwei Yu}
\orcid{0000-0002-2449-1436}
\affiliation{
  \institution{Microsoft Corporation}
  \city{Vancouver}
  \country{Canada}
}

\author{Lingwei Meng}
\orcid{0000-0003-1028-6017}
\affiliation{
  \institution{Microsoft Corporation}
  \city{Beijing}
  \country{China}
}

\author{Haiyang Sun}
\orcid{0009-0004-3485-3869}
\affiliation{
  \institution{Microsoft Corporation}
  \city{Beijing}
  \country{China}
}

\author{Yanqing Liu}
\orcid{0000-0002-4150-0680}
\affiliation{
  \institution{Microsoft Corporation}
  \city{Beijing}
  \country{China}
}

\author{Yan Lu}
\orcid{0000-0001-5383-6424}
\affiliation{
  \institution{Microsoft Corporation}
  \city{Beijing}
  \country{China}
}

\author{Kai Yu}
\orcid{0000-0002-7102-9826}
\affiliation{
  \institution{Shanghai Jiao Tong University}
  \city{Shanghai}
  \country{China}
}

\author{Xie Chen}
\orcid{0000-0001-7423-617X}
\affiliation{
  \institution{Shanghai Jiao Tong University, Shanghai Innovation Institute}
  \city{Shanghai}
  \country{China}
}

\renewcommand{\shortauthors}{Yifan Yang et al.}

\begin{abstract}
Recent zero-shot text-to-speech (TTS) systems face a common dilemma: autoregressive (AR) models suffer from slow generation and lack duration controllability, while non-autoregressive (NAR) models lack temporal modeling and typically require complex designs.
In this paper, we introduce a novel pseudo-autoregressive (PAR) codec language modeling approach that unifies AR and NAR modeling. Combining explicit temporal modeling from AR with parallel generation from NAR, PAR generates dynamic-length spans at fixed time steps.
Building on PAR, we propose PALLE, a two-stage TTS system that leverages PAR for initial generation followed by NAR refinement. 
In the first stage, PAR progressively generates speech tokens along the time dimension, with each step predicting all positions in parallel but only retaining the left-most span.
In the second stage, low-confidence tokens are iteratively refined in parallel, leveraging the global contextual information.
Experiments demonstrate that PALLE, trained on LibriTTS, outperforms state-of-the-art systems trained on large-scale data, including F5-TTS, E2-TTS, and MaskGCT, on the LibriSpeech test-clean set in terms of speech quality, speaker similarity, and intelligibility, while achieving up to ten times faster inference speed.
Audio samples are available at \url{https://microsoft.com/research/project/vall-e-x/palle}.
\end{abstract}



\maketitle

\begin{figure}[t!]
\centering
\includegraphics[width=0.972\linewidth]{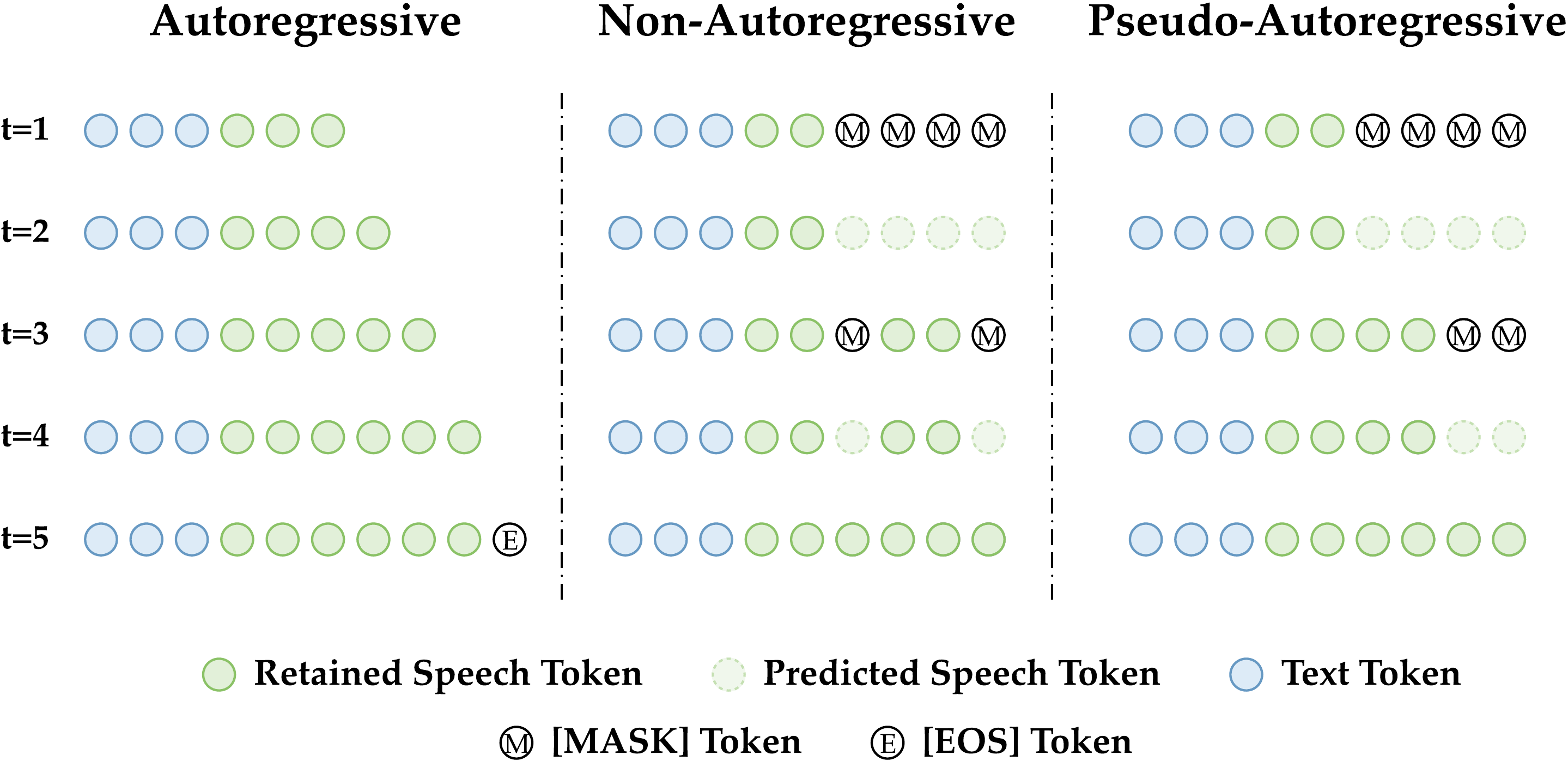}
\vspace{-0.5em}
\caption{Generation procedures of AR, NAR, and PAR.}
\Description{A diagram showing generation procedures across AR, NAR, and PAR paradigms.}
\label{fig:par}
\vspace{-1.5em}
\end{figure}

\section{Introduction}
Text-to-speech (TTS) synthesis has made remarkable progress, driven by generative methods~\cite{tacotron2, transformertts, vits, fastspeech2, difftts, valle} and the expansion of available data~\cite{didispeech, wenetspeech4tts, gigaspeech, gigaspeech2, libriheavy, emilia}.
These innovations have enabled TTS systems to achieve human-level parity in terms of naturalness and intelligibility, in both pre-defined specific speakers~\cite{naturalspeech} and zero-shot scenarios~\cite{valle2}.

Modern zero-shot TTS models can be broadly divided into two paradigms: autoregressive (AR)~\cite{valle, valle2, melle, felle, istlm, smlle, streammel, cosyvoice, cosyvoice2, cosyvoice3, voicecraft, basetts, seedtts, speartts, clamtts, ardit, ditar, syncspeech, fireredtts, indextts, indextts2, minimaxspeech, sled}\footnote{Hybrid systems like VALL-E are categorized by primary token generation.} and non-autoregressive (NAR)~\cite{naturalspeech2, naturalspeech3, voicebox, e2tts, e1tts, e3tts, f5tts, maskgct, yourtts, megatts, megatts2, megatts3, dittotts, zipvoice}.
AR models sequentially generate speech from left to right, conditioning each step on previous outputs.
Despite strong temporal grounding, AR models face fundamental limitations in generation efficiency and contextual modeling. Specifically, the fixed prediction size per step results in slow inference, with generation steps tightly coupled to the frame rate of output units. Additionally, the strictly unidirectional modeling restricts access to left context only and lacks future or global information, leading to exposure bias and posing challenges to robustness~\cite{ellav, valler, vallt}.
Techniques like grouped modeling~\cite{valle2} and speculative decoding~\cite{speculative_tts} improve efficiency, but generation steps still scale linearly with synthesized speech duration.
In contrast, NAR models enable fast inference through parallel generation. NAR TTS methods leverage denoising diffusion~\cite{naturalspeech3}, flow matching~\cite{e2tts, f5tts}, GAN~\cite{megatts2}, and masked generative modeling~\cite{maskgct}, drawing inspiration from advances in image~\cite{maskgit} and language~\cite{bert} generation.
Recent advances~\cite{e2tts, maskgct} demonstrate that text-to-speech alignment can be achieved without phoneme-level duration prediction or explicit alignment supervision.
E2-TTS~\cite{e2tts} and F5-TTS~\cite{f5tts} simply pad the character sequence with filler tokens to match the linearly predicted or user-specified speech length.
MaskGCT~\cite{maskgct} concatenates the tokenized text and speech sequences, adopts mask-and-predict training, and performs iterative parallel decoding during inference.
However, despite their efficiency and simplicity, these methods lack temporal modeling and generate speech in a temporally unordered manner, resulting in slow convergence~\cite{e2tts, f5tts}, alignment inaccuracies~\cite{e2tts}, and degraded intelligibility~\cite{maskgct}.

To combine the strengths of AR and NAR paradigms while mitigating their limitations, we propose a novel pseudo-autoregressive (PAR) codec language modeling paradigm that unifies AR and NAR modeling by combining temporal modeling from AR with parallel generation from NAR.
PAR introduces a soft temporal inductive bias into an NAR structure, specifically a bidirectional masked generative transformer, by enforcing span-level causal ordering through progressive commitment. 
Local spans are generated simultaneously, with earlier ones committed before later ones, creating a span-level temporal flow that aligns with the structure of natural speech.
Unlike grouped AR methods~\cite{valle2}, which generate fixed-length spans with dynamic time steps, PAR generates dynamic-length spans with fixed time steps.
Building on PAR, we propose \ours{}, a two-stage TTS system that performs PAR generation followed by confidence-guided refinement.
In the first stage, \ours{} progressively commits spans along the temporal dimension, generating a fixed portion of the target sequence at each step, resulting in constant inference steps regardless of target speech duration.
In the second stage, low-confidence tokens are refined in parallel based on global context, requiring only a few steps to correct accumulated errors and yielding huge performance improvements.

Experiments demonstrate that \ours{}, trained on LibriTTS, outperforms state-of-the-art systems trained on large-scale data, including F5-TTS, E2-TTS, and MaskGCT, on the LibriSpeech test-clean set in terms of speech quality, speaker similarity, and intelligibility, with inference speed up to 10$\times$ faster.

Our contributions can be summarized as:
\begin{itemize}[leftmargin=*, itemsep=0pt, topsep=0pt]
\item We propose and formulate a novel PAR codec language modeling paradigm for zero-shot TTS, which unifies AR and NAR paradigms by combining temporal modeling with parallel generation, offering new insights into generation algorithm design.
\item We instantiate AR, NAR, and PAR modeling paradigms under controlled conditions, and provide both empirical and theoretical indications of the effectiveness of PAR language modeling.
\item We introduce a new zero-shot TTS framework, namely \ours{}, which employs PAR for initial generation followed by NAR confidence-guided refinement.
\item We achieve a breakthrough in zero-shot TTS performance and efficiency, with the small-scale trained \ours{} outperforming large-scale state-of-the-art systems while enabling up to 10$\times$ faster inference.
\end{itemize}
\begin{figure*}[t!]
\centering
\includegraphics[width=\linewidth]{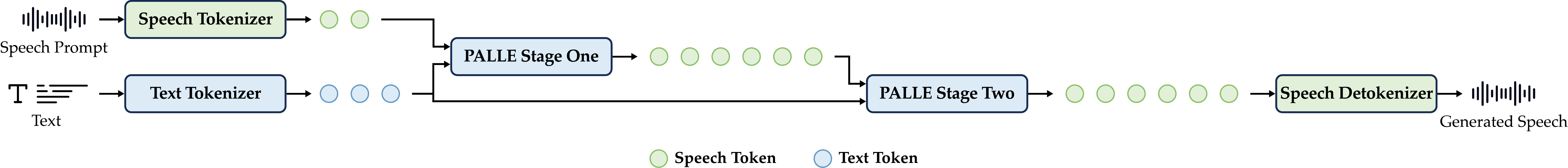}
\caption{
Overview of \ours{}, a two-stage codec language model for zero-shot TTS, comprising the following components: (1) a text tokenizer that converts text into subword tokens; (2) a speech tokenizer that encodes speech into a sequence of discrete speech tokens; (3) two masked generative transformer with shared architecture, where the first maps text tokens to speech tokens and the second refines generated speech tokens; and (4) a speech detokenizer with a built-in vocoder that converts the speech token sequence into waveform.
}
\Description{An overview diagram of \ours{}.}
\label{fig:palle_overview}
\vspace{-1em}
\end{figure*}

\section{Related work}
\subsection{Zero-Shot TTS}
Zero-shot TTS enables speech synthesis for unseen speakers by capturing the timbre, prosody, and style from short speech prompts.
Benefiting from in-context learning (ICL), the performance of zero-shot TTS has improved significantly, reaching human-parity quality~\cite{naturalspeech3, valle2}.
Recent approaches primarily fall into three categories: codec language modeling~\cite{valle, valle2, maskgct}, diffusion/flow-based methods~\cite{naturalspeech2, naturalspeech3, voicebox, e2tts, f5tts}, and hybrid approaches~\cite{cosyvoice, cosyvoice2}.
Hybrid systems involve a text-to-codec language model with a codec-to-mel model based on diffusion or flow matching, disentangling complex speech components to improve zero-shot generalization.

\subsection{Discrete Speech Representation}
Discrete speech representations~\cite{discretespeech} are typically categorized into acoustic tokens and semantic tokens.
Acoustic tokens, produced by neural codec models~\cite{encodec, dac, soundstream, wavtokenizer, stablecodec, ts3codec, funcodec}, aim to reconstruct speech with high fidelity, capturing fine-grained acoustic details.  
In contrast, semantic tokens are quantized from embeddings of self-supervised learning (SSL) models~\cite{vq-wav2vec, wav2vec2, hubert, wavlm}, which are trained for discrimination or masked prediction, and preserve the linguistic content of speech.
Recent methods have explored supervised semantic tokens~\cite{cosyvoice, cosyvoice2}, trained with automatic speech recognition (ASR) loss~\cite{ctc, rnnt}, making them more robust to noise and better suited for language modeling.

\subsection{Neural Codec Language Modeling}
Neural codec language modeling has emerged as a powerful paradigm for generative speech modeling.
Built upon discrete speech representations, language modeling approaches are employed to predict speech tokens, which are then decoded into waveforms by a speech detokenizer.
Two main paradigms have been explored: AR and NAR modeling.
AR models~\cite{valle, valle2, istlm, cosyvoice, cosyvoice2, speartts, basetts, syncspeech} generate tokens sequentially from left to right, providing strong temporal grounding but suffering from slow inference and limited context modeling. 
SyncSpeech~\cite{syncspeech} is an AR masked generative model that incorporates temporal modeling via a causal attention mask and token-level duration prediction, aiming for streaming generation from partial text input.
Both Grouped AR and SyncSpeech can predict multiple tokens at one time step, but the number of generation steps scales linearly with the input text, and only the previous context is considered during inference. In contrast, PAR generates the output sequence with fixed time steps while considering the global context.
NAR masked generative models~\cite{maskgct} improve generation efficiency by iteratively infilling masked positions using bidirectional context.
MaskGCT~\cite{maskgct} predicts the total speech length and performs temporally unordered prediction, relying solely on confidence-based scheduling, which may cause robustness issues in high-entropy regions such as fast transitions or expressive prosody.

\begin{figure*}[t!]
\centering
\includegraphics[width=0.95\linewidth]{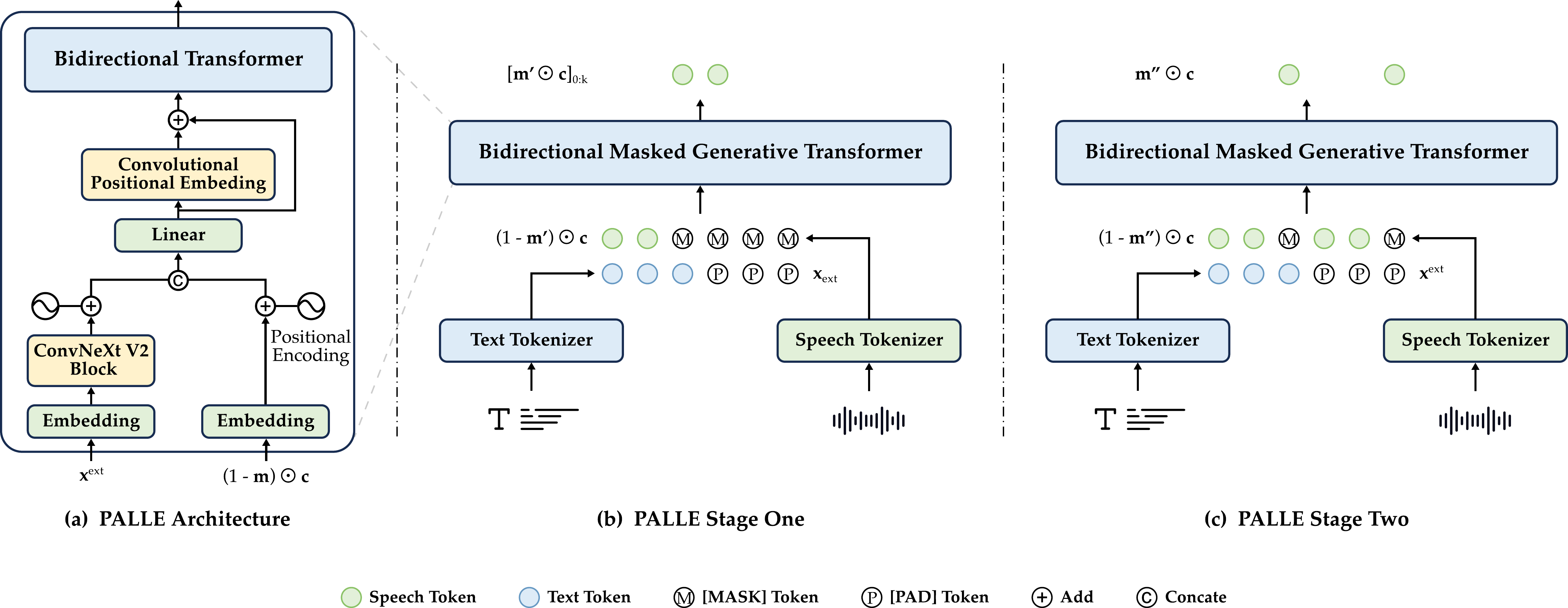}
\caption{
Illustration of the two-stage \ours{} framework.
(a) The shared architecture: a bidirectional masked generative transformer.
(b) Stage one: the PAR model predicts all token positions in parallel but retains only the leftmost span at each step.
(c) Stage two: the NAR model refines the initial speech tokens, where low-confidence tokens are re-masked and re-generated using contextual information.
}
\Description{An illustration diagram of \ours{} framework.}
\label{fig:palle_detail}
\end{figure*}

\section{Pseudo-Autoregressive Language Modeling}
We propose a pseudo-autoregressive (PAR) codec language modeling approach for zero-shot TTS that unifies NAR and AR modeling.
As illustrated in Fig.~\ref{fig:par}, PAR combines the parallel prediction from NAR with the temporal modeling from AR, enabling structured parallelism that follows the nature temporal order of speech.

\noindent\textbf{Formulation}\quad
Given a speech sample $\mathbf{y}$ and its corresponding tokenized text transcription $\mathbf{x}$, a pretrained speech tokenizer is used to encode the speech sample into discrete speech tokens $\mathbf{c}$ of downsampled length $T$, denoted as $\mathbf{c} = \mathrm{Encode}_\text{spch}(\mathbf{y})$.

The PAR model is built upon a bidirectional masked generative transformer~\cite{maskgit}.
During training, the model is optimized to maximize the likelihood of the speech tokens $\mathbf{c}$ conditioned on the text content $\mathbf{x}$ using cross-entropy loss. At each step, a contiguous span mask is applied to $\mathbf{c}$ from a random starting position to the end:
\begin{equation}
\label{eq:binary_mask}
\mathbf{m} = [\underbrace{0, 0, \ldots, 0}_{T - \ell}, \underbrace{1, 1, \ldots, 1}_{\ell}],
\end{equation}
where $\ell$ denotes the length of the masked span.
The model is optimized to predict the leftmost span of the masked portion $\mathbf{m} \odot \mathbf{c}$ with length $k = \lfloor r T \rfloor$, where $r \in (0, 1)$ is a fixed ratio. This prediction is conditioned on the unmasked portion $(1 - \mathbf{m}) \odot \mathbf{c}$ and the text content $\mathbf{x}$, corresponding to maximizing the following objective:
\begin{equation}
\arg\max_{\theta} \; p\left( [ \mathbf{m} \odot \mathbf{c}]_{0:k} \mid (1 - \mathbf{m}) \odot \mathbf{c}, \mathbf{x}; \theta \right),
\end{equation}
where $[\,\cdot\,]_{0:k}$ denotes the first $k$ elements, and $\theta$ represents the model parameters. The mask starting position is dynamically sampled during training, enabling the model to predict the next span along the time dimension based on varying amounts of preceding speech context. 
Note that all self-attention layers use full attention.

During inference, given a tokenized text $\mathbf{x}^\text{gen}$ to be synthesized, a speech prompt $\mathbf{y}^\text{ref}$ from an unseen speaker, and its corresponding tokenized text $\mathbf{x}^\text{ref}$, the speech tokens $\mathbf{c}^\text{ref} = [c'_0, c'_1, \ldots, c'_{T^\text{ref}-1}]$ of length $T^\text{ref}$ are extracted. The target speech token length $T^\text{gen}$ can be determined arbitrarily or estimated.
Inference begins by initializing an extended speech token sequence:
\begin{equation}
\mathbf{c}^\text{ext}_{(0)} = [c'_0, c'_1, \ldots, c'_{T^\text{ref}-1}, \underbrace{0, 0, \ldots, 0}_{T^\text{gen} - T^\text{ref}}],
\end{equation}
where the appended zeros serve as placeholders for the speech tokens to be generated.
At each step $t$, the model predicts all tokens in parallel, but retains only the leftmost $k' = \min(\lfloor r' T^\text{gen} \rfloor, N_\text{left})$ tokens, where $r' \in (0, 1)$ is a fixed ratio and $N_\text{left}$ is the number of tokens yet to be generated. Predictions are conditioned on the text prompt $\mathbf{x}^{\text{ref}}$, the target text $\mathbf{x}^{\text{gen}}$, and the current extended speech token sequence $\mathbf{c}^\text{ext}_{(t)}$.
A binary mask is constructed as follows:
\begin{equation}
\label{eq:binary_mask_iter}
\mathbf{m}_{(t)} = [\underbrace{0, 0, \ldots, 0}_{T^\text{ref} \; + \; t \times k'}, \underbrace{1, 1, \ldots, 1}_{k'}, 0, 0, \ldots, 0],
\end{equation}
The extended speech token sequence is updated according to the following iterative rule:
\begin{align}
\label{eq:par_iter_update}
\mathbf{c}^\text{ext}_{(t+1)} & = (1 - \mathbf{m}_{(t)}) \odot \mathbf{c}^\text{ext}_{(t)} \notag \\
& + \mathbf{m}_{(t)} \odot \arg\max_{\mathbf{c}^\text{ext}_{(t+1)}} p\left(\mathbf{c}^\text{ext}_{(t+1)} \mid \mathbf{c}^\text{ext}_{(t)}, \mathbf{x}^\text{ref}, \mathbf{x}^\text{gen}; \theta \right),
\end{align}
where the first part contains the tokens of the speech prompt and previously generated spans, and the second part contains the currently generated tokens.
This generation and selectively updating continue until all the tokens are generated.
The final speech tokens $\mathbf{c}^\text{ext}_{(\infty)}$ are converted into waveform $\hat{\mathbf{y}}$ using a pretrained speech detokenizer and vocoder, jointly denoted as $\hat{\mathbf{y}} = \mathrm{Decode}_\text{spch}(\mathbf{c}^\text{ext}_{(\infty)})$.

\noindent\textbf{Discussion}\quad
PAR is defined as a general language modeling paradigm that does not impose constraints on implementation details such as the fusion of text and speech tokens. It accommodates both temporal concatenation, as in MaskGCT~\cite{maskgct}, and feature-dimension fusion, where text and speech tokens are first padded to the same length, then independently embedded and concatenated along the feature dimension, as in E2 TTS~\cite{e2tts}.
\section{\ours{}}

\subsection{Architecture}
As illustrated in Fig.~\ref{fig:palle_overview}, given the text to be synthesized, a BPE-based text tokenizer converts the text into subword tokens, which are then processed by two masked generative transformers with a shared architecture. The first stage generates speech tokens with a PAR model, and the second stage refines the speech tokens with an NAR model. Finally, a speech detokenizer with a built-in vocoder converts the refined speech tokens into the waveform.

\noindent\textbf{Masked Generative Language Models}\quad
As depicted in Fig.~\ref{fig:palle_detail}(a), both stages in \ours{} adopt a bidirectional masked generative transformer~\cite{maskgit} as the backbone. The model processes two input sequences of equal length, a padded tokenized text sequence and a masked speech token sequence. 
The text embeddings are further processed by a ConvNeXt V2 block~\cite{convnextv2}, which provides strong temporal modeling capabilities, making the text representation better aligned with the speech sequence. A sinusoidal positional encoding is added for both sequences with a learnable scaling factor and dropout. The two sequences are concatenated along the feature dimension and passed through a linear projection layer. A convolutional positional embedding~\cite{voicebox} is then added to the projected features to encode temporal structure. The resulting sequence is fed into a bidirectional transformer that attends to the full context.

\noindent\textbf{Speech Tokenizer and Detokenizer}\quad
We use the pretrained S3Tokenizer from CosyVoice 2~\cite{cosyvoice} to extract discrete semantic tokens from the input waveform at a rate of 25 Hz. This model is a fine-tuned version of the SenseVoice-Large~\cite{funaudiollm} ASR model, trained on a large multilingual speech dataset for robust speech understanding. By leveraging ASR loss during training, the S3Tokenizer extracts semantic information, enabling implicit denoising and speaker disentanglement~\cite{touchtts}.
We adopt the pretrained conditional flow matching (CFM) model from CosyVoice 2~\cite{cosyvoice2} to reconstruct mel spectrograms from discrete speech tokens. 
The generated mel spectrogram is then converted into the waveform using the pretrained HiFi-GAN~\cite{hifi-gan} vocoder from CosyVoice 2~\cite{cosyvoice2}.

\subsection{Training: Conditional Codec Masked Language Modeling}
For simplicity, we reuse the symbol $\mathbf{c} = [c_0, c_1, \ldots, c_{T-1}]$ to denote a sequence of speech tokens obtained from a speech sample, and $\mathbf{x} = [x_0, x_1, \ldots, x_{L-1}]$ of length $L$ to represent its tokenized transcript.

\noindent\textbf{Stage One}\quad
The first stage of \ours{} implements PAR with the modality fusion method used in E2 TTS~\cite{e2tts}. As depicted in Fig.~\ref{fig:palle_detail}(b), the text token sequence $\mathbf{x}$ is padded with the filler token \texttt{[PAD]} to match the length of the speech token sequence $\mathbf{c}$, enabling fusion in the feature dimension after embedding. The padded text sequence $\mathbf{x}^\text{ext}$ is defined as:
\begin{equation}
\mathbf{x}^\text{ext} = [x_0, x_1, \ldots, x_{L-1}, \underbrace{\texttt{[PAD]}, \texttt{[PAD]}, \ldots, \texttt{[PAD]}}_{T - L}].
\end{equation}
At each step, a contiguous span of speech tokens in $\mathbf{c}$ is randomly masked using the binary mask $\mathbf{m'}$, which extends Eq.~\ref{eq:binary_mask} by keeping the first 30\% speech tokens unmasked as a prompt. The starting position $s$ is sampled from a uniform distribution:
\begin{equation}
s \sim \mathcal{U}\left\{ \lfloor 0.3T \rfloor, \lfloor 0.3T \rfloor + 1, \dots, T - \lfloor 0.1T \rfloor - 1 \right\},
\end{equation}
where the upper bound of the range ensures that at least the final 10\% of the sequence could be used for training.
Based on $s$, the binary mask $\mathbf{m'}$ is defined as:
\begin{equation}
\mathbf{m'} = [\underbrace{0, 0, \ldots, 0}_{s}, \underbrace{1, 1, \ldots, 1}_{T - s}].
\end{equation}
The model is optimized to predict the leftmost span of size $k = \left\lfloor 0.1 T \right\rfloor$ from the masked speech token sequence, denoted as $[\mathbf{m'} \odot \mathbf{c}]_{0:k}$, conditioned on the unmasked portion $(1 - \mathbf{m'}) \odot \mathbf{c}$ and the padded text sequence $\mathbf{x}^\text{ext}$.
The training objective is:  
\begin{equation}
\arg\max_{\theta} \; p\left( [ \mathbf{m'} \odot \mathbf{c}]_{0:k} \mid (1 - \mathbf{m'}) \odot \mathbf{c}, \mathbf{x}^\text{ext}; \theta \right).
\end{equation}

\noindent\textbf{Stage Two}\quad
Similar to stage one, $\mathbf{x}$ is padded with the filler token \texttt{[PAD]} to match the length of the speech token sequence $\mathbf{c}$.  
As illustrated in Fig.~\ref{fig:palle_detail}(c), speech tokens beyond $\lfloor 0.3 T \rfloor$ are independently masked with probability $p = 0.1$, forming a binary mask $\mathbf{m''} =\{0\}^{\lfloor 0.3 T \rfloor} \oplus \{0,1\}^{T-\lfloor 0.3 T \rfloor}$, where $\oplus$ means cancatenation and 1 indicating masked positions. The model is optimized to predict the masked speech tokens $\mathbf{m''} \odot \mathbf{c}$, conditioned on the unmasked speech tokens $(1 - \mathbf{m''}) \odot \mathbf{c}$ and the padded text sequence $\mathbf{x}^\text{ext}$, maximizing the following objective:  
\begin{equation}
\arg\max_{\theta} \; p\left( \mathbf{m''} \odot \mathbf{c} \mid (1 - \mathbf{m''}) \odot \mathbf{c}, \mathbf{x}^\text{ext}; \theta \right).
\end{equation}

\subsection{Inference: In-Context Learning Via Prompting}
We reuse the symbol $\mathbf{x}^\text{gen} = [x^g_0, x^g_1, \ldots, x^g_{L^\text{gen}-1}]$ of length $L^\text{gen}$ to represent the tokenized text to be synthesized, $\mathbf{y}^\text{ref}$ as the speech prompt, $\mathbf{x}^\text{ref} = [x^r_0, x^r_1, \ldots, x^r_{L^\text{ref}-1}]$ of length $L^\text{ref}$ as the tokenized text prompt, and $\mathbf{c}^\text{ref} = [c^r_0, c^r_1, \ldots, c^r_{T^\text{ref}-1}]$ of length $T^\text{ref}$ as the corresponding speech tokens from $\mathbf{y}^\text{ref}$.

\noindent\textbf{Stage One}\quad  
The target speech token length $T^\text{gen}$ is either predefined or estimated linearly as in \cite{f5tts}:
\begin{equation}
T^\text{gen} = T^\text{ref} \times ( 1 + \frac{L^\text{gen}}{L^\text{ref}}).
\end{equation}
The prompt speech token sequence $\mathbf{c}^\text{ref}$ is padded with the masked token \texttt{[MASK]} to match $T^\text{gen}$:  
\begin{equation}
\mathbf{c}^\text{ext} = [c^r_0, c^r_1, \ldots, c^r_{T^\text{ref}-1}, 
\underbrace{\texttt{[MASK]}, \texttt{[MASK]}, \ldots, \texttt{[MASK]}}_{T^\text{gen} - T^\text{ref}}].
\end{equation}
Meanwhile, the tokenized text sequences $\mathbf{x}^\text{ref}$ and $\mathbf{x}^\text{gen}$ are concatenated and padded with the filler token \texttt{[PAD]} to reach $T^\text{gen}$:  
\begin{equation}
\mathbf{x}^\text{ext} = [x^r_0, \ldots, x^r_{L^\text{ref}-1}, x^g_0, \ldots,  x^g_{L^\text{gen}-1}, 
\underbrace{\texttt{[PAD]}, \ldots, \texttt{[PAD]}}_{T^\text{gen} - (L^\text{ref} + L^\text{gen})}].
\end{equation}
The model progressively generates speech tokens along the time dimension. At each step $t$, the model predicts all tokens in parallel, retaining only the leftmost span of size $k'=  \min(\left\lfloor r' T \right\rfloor, N_\text{left})$, where $r' \in (0, 1)$ is a fixed ratio, and $N_\text{left}$ is the number of tokens yet to be generated
. Predictions are conditioned on the current extended speech and text token sequences, $\mathbf{c}^\text{ext}_{(t)}$ and $\mathbf{x}^\text{ext}$.
A binary mask $\mathbf{m}_{(t)}$ is constructed as in Eq.~\ref{eq:binary_mask_iter}, and the extended speech token sequence is updated iteratively, as in Eq.\ref{eq:par_iter_update}:
\begin{align}
\label{eq:ext_iter}
\mathbf{c}^\text{ext}_{(t+1)} & = (1 - \mathbf{m}_{(t)}) \odot \mathbf{c}^\text{ext}_{(t)} \notag \\
& + \mathbf{m}_{(t)} \odot \arg\max_{\mathbf{c}^\text{ext}_{(t+1)}} p\left(\mathbf{c}^\text{ext}_{(t+1)} \mid \mathbf{c}^\text{ext}_{(t)}, \mathbf{x}^\text{ext}; \theta \right).
\end{align}

\noindent\textbf{Stage Two}\quad
In this stage, the model iteratively refines the initial generation $\mathbf{c'}^\text{ext}_{(0)} = \mathbf{c}^\text{ext}_{(\infty)}$ from stage one, by re-masking and re-predicting low-confidence tokens.  
At each step $t$, the model produces a probability matrix $\mathbf{P}_{(t)}$, where each entry $P_{(t),n}$ denotes the predicted probability of class $n$ out of $N$ speech token classes, conditioned on $\mathbf{c'}^\text{ext}_{(t)}$ and $\mathbf{x}^\text{ext}$:
\begin{equation}
\mathbf{P}_{(t)} = p\left(\mathbf{c'}^\text{ext}_{(t+1)} \mid \mathbf{c'}^\text{ext}_{(t)}, \mathbf{x}^\text{ext}; \theta \right) \in \mathbb{R}^{T^\text{gen} \times N}.
\end{equation}
The confidence score matrix $\mathbf{C}_{(t)} \in \mathbb{R}^{T^\text{gen}}$ is defined as the negative min-entropy of the corresponding distribution:
\begin{equation}
\mathbf{C}_{(t)} = \log \mathbf{P}^\text{max}_{(t)}[:, n],
\end{equation}
where $\mathbf{P}^\text{max}_{(t)}$ is the maximum probabilities in the predicted distributions.
We rank the confidence scores and select those tokens with the lowest confidence for re-masking using a predefined quantile $\gamma$. 
A binary mask $\mathbf{m}_{(t)} \in \{0,1\}^{T^\text{gen}}$ is constructed, where 1 marks the selected positions for masking.  
The extended speech token sequence $\mathbf{c'}^\text{ext}_{(t)}$ is updated iteratively as in Eq.~\ref{eq:ext_iter}.
To prevent repeated refinement at the same position, the confidence scores of updated tokens are set to 1 permanently.
The final speech tokens $\mathbf{c'}^\text{ext}_{(\infty)}$ are converted into waveform $\hat{\mathbf{y}}$.

\begin{table*}[t]
\small
\centering
\caption{
Objective performance comparison on continuation and cross-sentence tasks.
\textbf{Bold} highlights the best result, while \underline{underlined} marks the second-best. $^*$Metrics not reported in the original papers are calculated using the open-source checkpoints. \textit{GT duration} uses ground-truth total speech duration. For \textit{Continuation} task, NAR systems without \textit{GT duration} are excluded, as linear duration estimation requires forced alignment to obtain the transcription of the first three seconds of speech.
}

\renewcommand\tabcolsep{2pt}
\resizebox{\linewidth}{!}{
\begin{tabular}{lccccccccc}
\toprule[1pt]
\multirow{2}{*}{\textbf{System}} & \multirow{2}{*}{\textbf{Speech Tokenizer}} & \multirow{2}{*}{\textbf{Training Data}} & \multicolumn{3}{c}{\textbf{Continuation}} & \multicolumn{4}{c}{\textbf{Cross-Sentence}} \\
\cmidrule(r){4-6} \cmidrule(r){7-10} & & & WER-W$\downarrow$ & WER-H$\downarrow$ & SIM-o$\uparrow$ & WER-W$\downarrow$ & WER-H$\downarrow$ & SIM-o$\uparrow$ & RTF$\downarrow$ \\
\midrule
Ground Truth                       & - & - & 2.14 & 2.15 & 0.905 & 2.14 & 2.15 & 0.779 & - \\
Ground Truth (EnCodec)             & - & - & -    & 2.33 & 0.823 & -    & 2.33 & 0.715 & - \\
Ground Truth (S3Tokenizer v2 25Hz) & - & - & 2.69 & 3.16 & 0.793 & 2.94 & 3.51 & 0.743 & - \\
\midrule
\textbf{Trained on Large-Scale Dataset} \\ 
VALL-E~\cite{valle} & EnCodec~\cite{encodec} & Librilight & - & 3.80 & \underline{0.773} & - & 5.90 & 0.633 & 0.73 \\
VALL-E ~\cite{valle_2025} & EnCodec~\cite{encodec} & Libriheavy & - & \underline{3.00} & 0.770 & - & 3.80 & 0.630 & 0.73 \\
E2 TTS (32 NFE)$^*$ & - & Emilia (EN + ZH) & - & - & - & 2.55 & 2.92 & \textbf{0.756} & 0.68 \\
F5-TTS (32 NFE)$^*$ & - & Emilia (EN + ZH) & - & - & - & \underline{2.25} & \textbf{2.77} & 0.705 & \underline{0.15} \\
MaskGCT$^*$ & Semantic Codec + Acoustic Codec~\cite{maskgct} & Emilia (EN + ZH) & - & - & - & 3.69 & 4.22 & \textbf{0.756} & 0.65 \\
\midrule
\textbf{Trained on Small-Scale Dataset} \\
VALL-E                                              & EnCodec~\cite{encodec}                & LibriTTS & \underline{3.78} & 4.47 & 0.730 & 7.36 & 8.64 & 0.531 & 0.73 \\ 			
CosyVoice~\cite{syncspeech}                         & S3Tokenizer v1 25Hz~\cite{cosyvoice}  & LibriTTS & - & - & - & 3.47 & - & - & 0.45 \\
CosyVoice 2~\cite{syncspeech}                       & S3Tokenizer v2 25Hz~\cite{cosyvoice2} & LibriTTS & - & - & - & 3.00 & - & - & 0.45 \\
\ours{}$_{\text{two-stage}}$                        & S3Tokenizer v2 25Hz~\cite{cosyvoice2} & LibriTTS & - & - & - & \textbf{2.23} & \underline{2.83} & \underline{0.716} & \textbf{0.06} \\
\ours{}$_{\text{two-stage}}$ \textit{(GT duration)} & S3Tokenizer v2 25Hz~\cite{cosyvoice2} & LibriTTS & \textbf{2.31} & \textbf{2.62} & \textbf{0.776} & 2.35 & 2.87 & \underline{0.716} & \textbf{0.06} \\
\bottomrule[1pt]
\end{tabular}
}
\label{tab:main_obj}
\end{table*}

\begin{table}[ht]
\small
\centering
\caption{
Subjective performance comparison on cross-sentence zero-shot speech synthesis tasks. \textit{GT duration} denotes the results obtained using the ground-truth total speech duration.
SMOS scores are accompanied by 95\% confidence intervals (CI) and p-values from Wilcoxon signed-rank tests for statistical significance.
}

\begin{tabular}{lccc}
\toprule[1pt]
\textbf{System} & SMOS$\uparrow$ & CMOS$\uparrow$ & p-value \\
\midrule
Ground Truth    & 4.23 $\pm$ 0.18 & 0 & - \\
\midrule
E2 TTS (32 NFE) & 3.95 $\pm$ 0.20 & -0.41 & 0.031 \\
F5-TTS (32 NFE) & 4.04 $\pm$ 0.18 & -0.20 & 0.033 \\
MaskGCT         & 3.98 $\pm$ 0.17 & -0.25 & 0.025 \\
\midrule
\ours{}$_{\text{two-stage}}$ & 4.03  $\pm$ 0.17 & -0.15 & 0.036 \\
\ours{}$_{\text{two-stage}}$ (\textit{GT duration}) & \textbf{4.09} $\pm$ 0.19 & \textbf{-0.09} & 0.038 \\
\bottomrule[1pt]
\end{tabular}
\label{tab:main_sub}
\end{table}
\begin{table*}[t]
\small
\centering
\caption{
Objective comparison of representative AR, NAR, and PAR models on continuation and cross-sentence zero-shot synthesis tasks.
All models are trained on LibriTTS using the same speech tokenizer, detokenizer, and Transformer backbone.
}

\begin{tabular}{llcccccccc}
\toprule[1pt]
\multirow{2}{*}{\textbf{ID}} & \multirow{2}{*}{\textbf{System}} & \multirow{2}{*}{\textbf{Language Modeling Paradigm}} & \multicolumn{3}{c}{\textbf{Continuation}} & \multicolumn{4}{c}{\textbf{Cross-Sentence}} \\
\cmidrule(r){4-6} \cmidrule(r){7-10} & & & WER-W$\downarrow$ & WER-H$\downarrow$ & SIM-o$\uparrow$ & WER-W$\downarrow$ & WER-H$\downarrow$ & SIM-o$\uparrow$ & RTF$\downarrow$ \\
\midrule
\textbf{A} & VALL-E$_\text{stage-one-only}$ & AR & 3.32 & 3.35 & \textbf{0.776} & 4.00 & 4.32 & \textbf{0.716} & 0.21 \\
\textbf{B} & MaskGCT$_\text{stage-one-only}$ & NAR & -    & -    & -     & 4.52 & 5.18 & 0.703 & 0.10 \\
\textbf{C1} & \ours{}$_{\text{stage-one-only}}$ & PAR & -    & -    & -     & 2.58 & 3.15 & 0.710 & \textbf{0.05} \\
\textbf{D1} & \ours{}$_{\text{stage-one-only}}$ \textit{(GT duration)} & PAR & 2.41 & 2.70 & 0.775 & 2.64 & 3.14 & 0.711 & \textbf{0.05} \\
\textbf{C2} & \ours{}$_{\text{two-stage}}$ & PAR + NAR & -    & -    & -     & \textbf{2.23} & \textbf{2.83} & \textbf{0.716} & 0.06 \\
\textbf{D2} & \ours{}$_{\text{two-stage}}$ \textit{(GT duration)} & PAR + NAR & \textbf{2.31} & \textbf{2.62} & \textbf{0.776} & 2.35 & 2.87 & \textbf{0.716} & 0.06 \\
\bottomrule[1pt]
\end{tabular}
\label{tab:par}
\end{table*}

\section{Experiments}

\subsection{Experimental Setup}
\noindent\textbf{Training Dataset}\quad
We conduct experiments on the LibriTTS~\cite{libritts} dataset, a widely used multi-speaker English corpus containing approximately 580 hours of speech from 2,306 speakers.

\noindent\textbf{Model Configurations}\quad
Both stages of \ours{} adopt the same model architecture, comprising: a decoder-only Transformer with 12 layers, 16 attention heads, 1024-dimensional embeddings, and 4096-dimensional feed-forward layers; a one-layer ConvNeXt V2 block with 1024-dimensional embeddings and a 2048-dimensional feed-forward layer; and a convolutional positional embedding with a kernel size of 7; totaling 177.0M parameters.
For text tokenization, a 2,000-class BPE is used. Speech tokenization is carried out using the off-the-shelf S3Tokenizer\footnote{\url{https://github.com/xingchensong/S3Tokenizer}} from CosyVoice 2~\cite{cosyvoice2} at a token rate of 25 Hz. Speech reconstruction is performed using the off-the-shelf CFM model with a built-in vocoder, also from CosyVoice 2~\cite{cosyvoice2}.

\noindent\textbf{Training Details}\quad
Both stages are trained on 8 NVIDIA TESLA V100 32GB GPUs, with a batch duration of 600 seconds per GPU. The first stage is trained for 179k steps, and the second stage is fine-tuned based on the first stage for another 87k steps. The ScaledAdam~\cite{zipformer} optimizer and Eden~\cite{zipformer} scheduler are used, with a peak learning rate of 0.045 for the first stage and 0.005 for the second stage.

\begin{figure*}[t]
\centering

\begin{subfigure}[t]{0.24\linewidth}
    \centering
    \includegraphics[width=\linewidth]{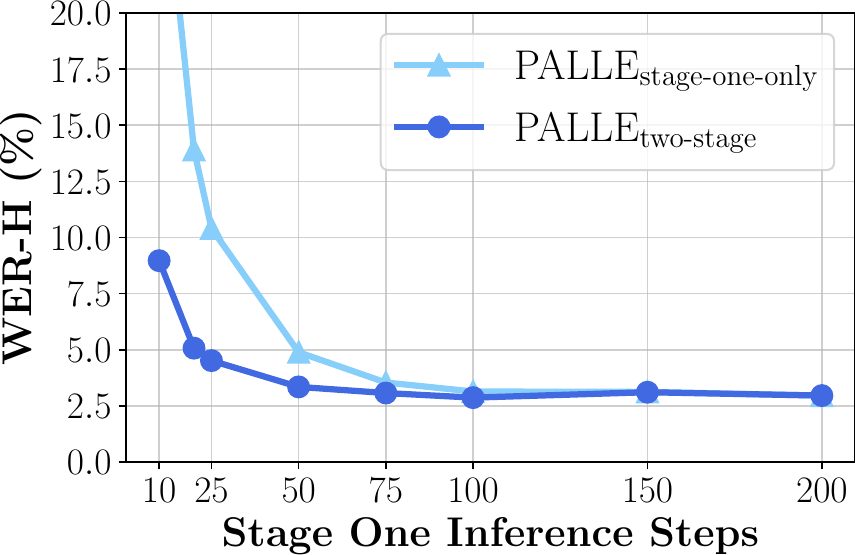}
    \label{fig:stage_one_infer_step_to_wer}
\end{subfigure}
\hfill
\begin{subfigure}[t]{0.24\linewidth}
    \centering
    \includegraphics[width=\linewidth]{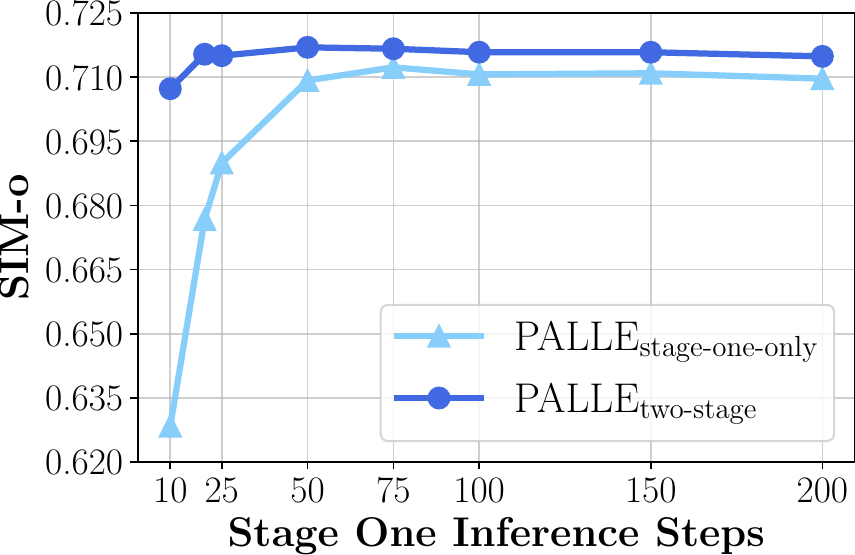}
    \label{fig:stage_one_infer_step_to_sim}
\end{subfigure}
\hfill
\begin{subfigure}[t]{0.24\linewidth}
    \centering
    \includegraphics[width=\linewidth]{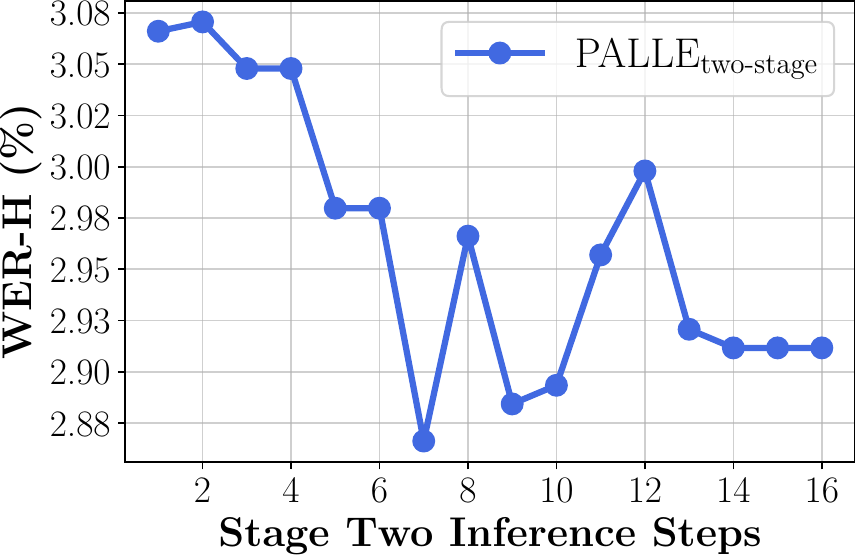}
    \label{fig:stage_two_infer_step_to_wer}
\end{subfigure}
\hfill
\begin{subfigure}[t]{0.24\linewidth}
    \centering
    \includegraphics[width=\linewidth]{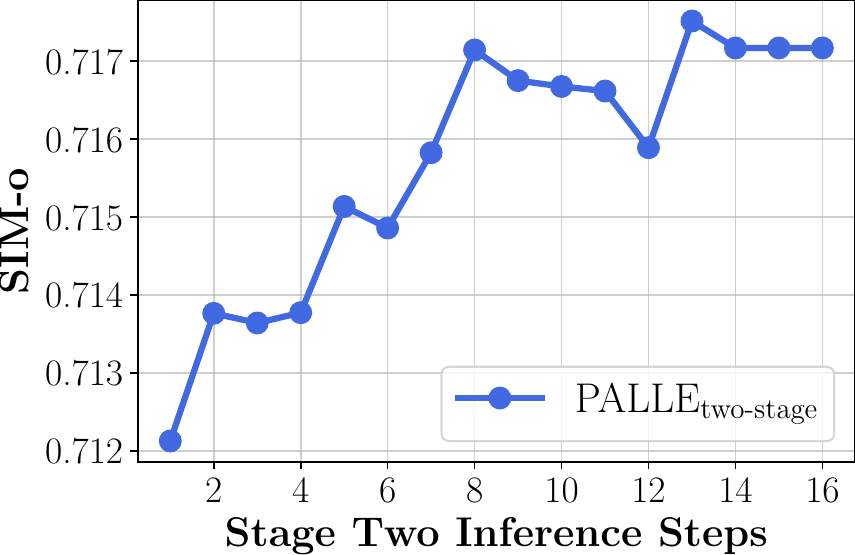}
    \label{fig:stage_two_infer_step_to_sim}
\end{subfigure}
\vspace{-1em}

\caption{
Impact of varying inference steps on objective performance metrics for cross-sentence zero-shot speech synthesis. From left to right: (a) Stage One Inference Steps on WER-H; (b) Stage One Inference Steps on SIM-o; (c) Stage Two Inference Steps on WER-H; (d) Stage Two Inference Steps on SIM-o.
}
\Description{Four side-by-side charts.}
\label{fig:infer_step_analysis}
\end{figure*}

\noindent\textbf{Inference Details}\quad
The first stage runs 100 steps per sample of length $T$, each generating a span of size $k = \min(\left\lfloor 0.01 T \right\rfloor, N_\text{left})$, which corresponds to a fixed percentage of the sample length and is smaller than the span size used during training\footnote{Although the model is trained with span size of $\left\lfloor 0.1 T \right\rfloor$, we found the best configuration for inference is $\left\lfloor 0.01 T \right\rfloor$, as detailed in Sec.~ \ref{sec:ablation}.}. A detailed analysis of the per-step prediction ratio is provided in Fig.~\ref{fig:infer_step_analysis} of the ablation study.
The second stage runs 7 steps per sample using the $\gamma = 0.05$ quantile of the confidence scores.
Top-$p$ sampling with a small $p$ in the range of 0.2 to 0.35 yields slight performance improvements over greedy search, without introducing much randomness or additional computational cost.

\subsection{Evaluation}
\noindent\textbf{Evaluation Settings}\quad
We use the LibriSpeech~\cite{librispeech} test-clean set for zero-shot TTS evaluation, ensuring no speaker overlap with training data. Following previous works~\cite{valle,valle_2025}, we use the same test set, which consists of segments ranging from 4 to 10 seconds, totaling 2.2 hours of data from 40 unique speakers and 1,234 samples.
Our evaluations contain two inference tasks:
\begin{itemize}[leftmargin=*, itemsep=0pt, topsep=2pt]
\item \textit{Continuation}: Given the text and the first 3 seconds of an utterance as the prompt, the model generates the remaining speech.
\item \textit{Cross-Sentence}: Given a reference utterance and its transcription as the prompt, the model synthesizes speech for the target text.
\end{itemize}

\noindent\textbf{Objective Metrics}\quad
We adopt the following objective metrics to assess the naturalness, robustness, speaker similarity, and efficiency of the proposed method.
For the continuation task, we evaluate the entire utterance rather than only the continuation segment for a more comprehensive comparison.
\begin{itemize}[leftmargin=*, itemsep=0pt, topsep=2pt]
\item \textbf{WER} (Word Error Rate) measures the robustness and intelligibility of synthesized speech.
WER is computed between ASR transcription of synthesized audio and the ground-truth text, denoted as WER-W and WER-H, with Whisper Large-v3\footnote{\url{https://huggingface.co/openai/whisper-large-v3}}~\cite{whisper} and the HuBERT-Large ASR model\footnote{\url{https://huggingface.co/facebook/hubert-large-ls960-ft}}~\cite{hubert} respectively.

\item \textbf{SIM-o} (Speaker Similarity) measures the speaker similarity between the original speech and the synthesized speech. The state-of-the-art speaker verification model WavLM-TDNN\footnote{\url{https://github.com/microsoft/UniSpeech/tree/main/downstreams/speaker_verification\#pre-trained-models}}~\cite{wavlm} is employed for evaluation. The similarity score ranges from $[-1, 1]$, with higher values indicating greater speaker similarity.

\item \textbf{RTF} (Real-Time Factor) measures the time taken to synthesize one second of speech and reflects system efficiency, especially in real-time scenarios. We report RTF on an NVIDIA TESLA A100 80G GPU, calculated from the average inference time for generating 10 seconds of speech with a batch size of 1.
\end{itemize}

\noindent\textbf{Subjective Metrics}\quad
We use the following subjective evaluation metrics to assess the speaker similarity and comparative quality of each synthesized speech.
Twenty native English speakers with experience in speech annotation and evaluation are engaged as contributors in a crowd-sourced evaluation.

\begin{itemize}[leftmargin=*, itemsep=0pt, topsep=2pt]
\item \textbf{SMOS} (Similarity Mean Opinion Score) for measuring speaker similarity between the speech prompt and the generated speech. The SMOS scale ranges from 1 to 5, in 0.5-point increments.

\item \textbf{CMOS} (Comparative Mean Opinion Score) for evaluating the comparative quality of the synthesized speech against the original ground-truth audio. The CMOS scale ranges from -3 (much worse than the ground truth) to 3 (much better than the ground truth), in 1-point increments.
\end{itemize}

\newpage

\noindent\textbf{Baseline Details}\quad
We evaluate our models against several state-of-the-art zero-shot TTS systems.
\begin{itemize}[leftmargin=*, itemsep=0pt, topsep=2pt]
\item \textbf{VALL-E}~\cite{valle}: A two-stage AR+NAR system predicting 75Hz EnCodec RVQ tokens~\cite{encodec}. We evaluate three variants trained on Librilight~\cite{librilight}, Libriheavy~\cite{libriheavy} and LibriTTS~\cite{libritts}, with RTF from~\cite{melle} using the official checkpoint.

\item \textbf{CosyVoice}~\cite{cosyvoice}: A two-stage AR+flow-matching system.
We evaluate the 25Hz LibriTTS version, with results from~\cite{syncspeech}.

\item \textbf{CosyVoice 2}~\cite{cosyvoice2}: A two-stage interleaved AR+chunk-aware flow-matching system. We evaluate the 25Hz LibriTTS version, with results from~\cite{syncspeech}.

\item \textbf{MaskGCT}~\cite{maskgct}: A two-stage NAR mask-and-predict system with 695M text-to-semantic and 353M semantic-to-acoustic models. We evaluate the official checkpoint\footnote{\url{https://huggingface.co/amphion/MaskGCT}} trained on 100K hours of Chinese and English speech from Emilia~\cite{emilia}.

\item \textbf{E2 TTS}~\cite{e2tts} \& \textbf{F5-TTS}~\cite{f5tts}: Fully NAR flow-matching systems with 333M and 336M parameters. We evaluate open-source checkpoints\footnote{\url{https://huggingface.co/SWivid/E2-TTS}}\footnote{\url{https://huggingface.co/SWivid/F5-TTS}} trained on 100K hours of bilingual Emilia~\cite{emilia}.
\end{itemize}

\begin{figure}[t!]
\centering

\begin{subfigure}[t]{0.48\linewidth}
    \centering
    \includegraphics[width=\linewidth]{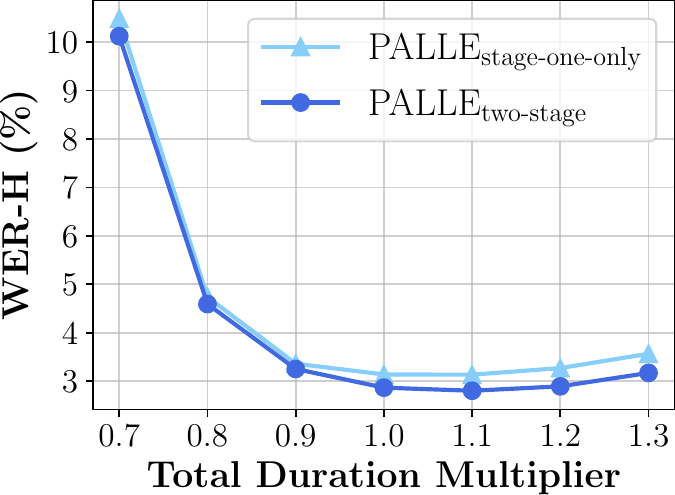}
    \label{fig:duration_to_wer}
\end{subfigure}
\hfill
\begin{subfigure}[t]{0.48\linewidth}
    \centering
    \includegraphics[width=\linewidth]{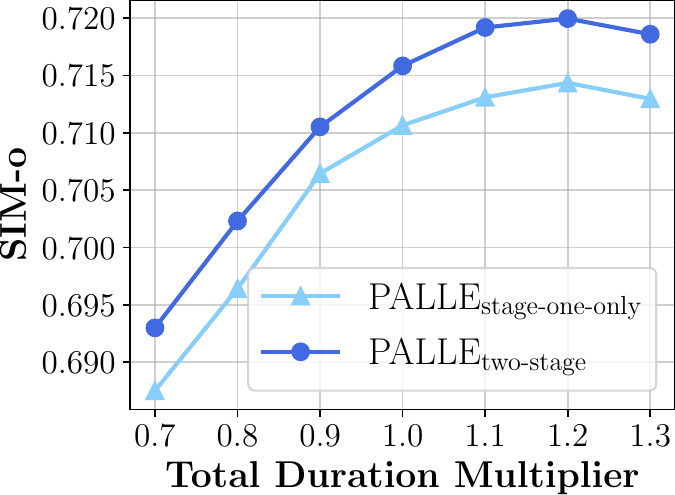}
    \label{fig:duration_to_sim}
\end{subfigure}

\vspace{-1em}

\caption{
Impact of varying the total duration on objective performance for cross-sentence zero-shot speech synthesis. The left plot shows the effect of the total duration multiplier on WER-H (\%), and the right plot shows the effect on SIM-o.
}
\Description{Two side-by-side charts. The left shows WER versus total duration multiplier. The right shows SIM versus total duration multiplier.}
\label{fig:duration_analysis}
\end{figure}

\subsection{Evaluation Results}

\noindent\textbf{Objective Evaluation}\quad
Table~\ref{tab:main_obj} presents comparisons between our proposed \ours{} and the baselines in terms of robustness, similarity, and efficiency on the LibriSpeech test-clean set, where \ours{} demonstrates superior performance on both continuation and cross-sentence tasks.
Notably, despite being trained on small-scale data, \ours{} outperforms or matches state-of-the-art systems trained on large-scale data, including F5-TTS, E2 TTS, and MaskGCT, and achieves up to 10$\times$ faster inference with an extremely low RTF of 0.06, demonstrating PAR as a powerful and efficient paradigm for codec language modeling and \ours{} as a practical solution for real-time scenarios.

On the continuation task, \ours{} with GT duration achieves the best WER and SIM among all systems. On the cross-sentence task, \ours{} with linearly estimated duration achieves the best WER and the second-best SIM.
Specifically, \ours{} significantly outperforms MaskGCT and E2 TTS in intelligibility, with a 40\% relative reduction in WER-W over MaskGCT, maintains competitive similarity with only a 5\% relative drop, and achieves over 10$\times$ higher efficiency.
Compared to the highly optimized F5-TTS, \ours{} matches its robustness, surpasses it in similarity, and is 2$\times$ faster in inference speed.
When compared to the LibriTTS-trained CosyVoice 2, which uses the same speech tokenizer and detokenizer, \ours{} with GT duration delivers a 26\% improvement in WER-W and is 7.5 times faster, both at a token rate of 25Hz.

Compared to ground-truth speech reconstructed by S3Tokenizer v2, \ours{} with GT duration yields a 17\% relative WER-H reduction on continuation, while \ours{} with estimated duration achieves a 24\% relative WER-W reduction on cross-sentence. In the latter case, it performs nearly on par with the ground-truth speech in WER-W.
We also observe that \ours{} with estimated duration slightly exceeds \ours{} with GT duration in intelligibility while maintaining comparable similarity, suggesting the effectiveness of our BPE-level linear duration estimation in producing appropriate durations.

\noindent\textbf{Subjective Evaluation}\quad
The subjective evaluation uses one utterance per speaker from the LibriSpeech test-clean set, totaling 40 samples. For each sample, the preceding utterance from the official list is used as the prompt to generate the target utterance.
Table~\ref{tab:main_sub} reports comparisons between \ours{} and the baselines in terms of perceived speaker similarity and comparative quality, where \ours{} with GT duration consistently outperforms all baselines across both metrics, producing speech that is virtually indistinguishable from the ground truth, with only a 0.09 gap in CMOS.
\ours{} with estimated duration achieves the highest CMOS, surpasses E2 TTS and MaskGCT, and performs on par with F5-TTS in SMOS.

\subsection{Pseudo-Autoregressive \textit{vs.} Autoregressive and Non-Autoregressive}
To further assess the effectiveness of the proposed PAR modeling paradigm, we conduct a comparative study with representative AR and NAR approaches under controlled conditions. As detailed in Table~\ref{tab:par}, we train a VALL-E stage-one AR model (Sys.~A)  and a MaskGCT stage-one NAR model (Sys.~B) on LibriTTS, using the same speech tokenizer, detokenizer, and Transformer backbone, to be compared with our PAR model (Sys.~C1).
Collectively, these three models serve as representative instantiations of the AR, NAR, and PAR modeling paradigms, forming a fair and isolated comparison.

\noindent\textbf{PAR \textit{vs.}~AR}\quad
Compared to the AR-based Sys.~A, our PAR-based Sys.~C1 achieves 36\% relative reduction in WER-W and 4$\times$ speedup in inference, while maintaining comparable speaker similarity.
These performance gains stem from the fundamental distinct modeling paradigms.
AR models generate tokens sequentially, resulting in inference latency $\mathcal{O}(T)$, where $T= \mu D$, tied to speech token rate $\mu$ and growing linearly with duration $D$.
In contrast, our PAR model generates a dynamic-length span of tokens in parallel at each step. Since each span covers a fixed proportion of the target sequence, the number of decoding steps remains constant, resulting in $\mathcal{O}(1)$ inference latency and substantial acceleration under the same token rate.
Regarding accuracy, AR models rely solely on left-context for each prediction, leading to error accumulation as the sequence length increases.
PAR mitigates this by predicting the entire sequence in parallel at each step and progressively committing only to the left-most span, enabling more robust and flexible temporal modeling with bidirectional context.

\noindent\textbf{PAR \textit{vs.}~NAR}\quad
On the cross-sentence task, our PAR-based Sys.~C1 yields a 43\% relative WER-W reduction, 2$\times$ faster inference, and marginally better speaker similarity compared to the NAR-based Sys.~B.
While both NAR and PAR models offer $\mathcal{O}(1)$ latency through parallel decoding, the temporally unordered generation in NAR leads to poor alignment, requiring more decoding steps $1 / r'$ with a smaller predicted proportion $r'$ per step to maintain output quality. This results in compromised robustness, degraded prosody, and reduced efficiency in practice.
PAR addresses this by introducing a soft temporal inductive bias through span-level causal ordering, preserving the natural temporal flow of speech and improving alignment and intelligibility.

\subsection{Ablation Study}
\label{sec:ablation}

\noindent\textbf{Total Duration}\quad
We evaluate the robustness of the stage-one and two-stage variants of \ours{} under varying total duration multipliers on the cross-sentence task, where each multiplier corresponds to a different speech tempo.  
As shown in Fig.~\ref{fig:duration_analysis}, the duration of the generated speech is scaled from 0.7$\times$ to 1.3$\times$ of the ground-truth duration.  
Both systems achieve the lowest WER-H at a multiplier of 1.1 and the highest SIM at 1.2, indicating that generating speech at a slightly slower pace than the ground truth improves both intelligibility and speaker similarity.  
One possible explanation is that, in the cross-sentence task, the speaking rate of the speech prompt may differ from that of the target. When the prompt is substantially faster, the model may struggle to maintain prosodic consistency via in-context learning, leading to degraded synthesis quality. 
Overall, both systems exhibit stable performance across a moderate tempo range, with near-optimal results between 0.9$\times$ and 1.3$\times$, indicating robustness to speaking rate variation.

\noindent\textbf{Inference Steps}\quad
Fig.~\ref{fig:infer_step_analysis} illustrates the impact of inference steps in both stage one and stage two. 
In the first two plots, we vary the number of steps in stage one while keeping stage two fixed. Both curves show rapid performance gains with more steps, followed by a plateau. We select 100 steps for stage one as the optimal point to balance efficiency and performance.
In the last two plots, we vary the number of steps in stage two. Although the curves exhibit some fluctuations, the overall trend improves with more steps and stabilizes after step 14. We select 7 steps for stage two as the optimal point.
Although stage two involves only a few steps, it brings clear improvements across both metrics compared to stage one.

\noindent\textbf{Joint Two-Stage Modeling}\quad  
We attempt to unify stage one and stage two into a single model via multitask finetuning, randomly performing one task per sample.  
However, stage two loss degrades stage one performance.  
While performance slightly decreases on the continuation task, WER on the cross-sentence task increases by 20\%, regardless of whether ground-truth durations are used. This may be attributed to the halved parameter capacity per stage compared to the separate models.
\section{Conclusions}
In this paper, we propose a novel sequence generation paradigm, pseudo-autoregressive language modeling (PAR), which combines the strengths of temporal modeling from AR and parallel decoding from NAR, and generates dynamic-length spans at fixed time steps. We apply PAR to zero-shot TTS tasks, resulting in a new TTS model called \ours{}. \ours{} leverages a PAR model to generate initial discrete speech tokens with fixed steps for various target utterances, followed by refinement using a NAR model. Experimental results demonstrate that \ours{} significantly improves intelligibility, speaker similarity, and efficiency. Looking ahead, PAR holds potential for broader applicability in generation domains, such as text, image, and video.

\clearpage

\begin{acks}
This work was supported by the National Natural Science Foundation of China (No. 62206171 and No. U23B2018), Shanghai Municipal Science and Technology Major Project under Grant 2021SHZDZX0102, and Yangtze River Delta Science and Technology Innovation Community Joint Research Project (2024CSJGG01100).
\end{acks}

\bibliographystyle{ACM-Reference-Format}
\balance
\bibliography{sample-base}

\end{document}